\documentstyle[12pt]{article}
\begin{document}

\title{On dipole compression modes in nuclei~}

\author{G. Col\`o$^a$\thanks{
        Tel.: + 39 - 02 - 2392261; fax: + 39 - 02 - 2392487; e-mail: colo@mi.infn.it}, 
        N. Van Giai$^b$, P.F. Bortignon$^a$ and
        M.R. Quaglia$^c$}
\maketitle

{\noindent $^a$ Dipartimento di Fisica, Universit\`a degli Studi,  
        and INFN, Via Celoria 16, I-20133 Milano (Italy) \\
        $^b$ 
	Institut de Physique 
        Nucl\'eaire, IN2P3-CNRS, F-91406 Orsay (France) \\
        $^c$ Dipartimento di Fisica Teorica, Universit\`a degli Studi,  
        and INFN, via P. Giuria 1, I-10125 Torino (Italy) }

\begin{abstract}
Isoscalar dipole strength distributions in spherical 
medium- and heavy-mass nuclei are
calculated within random phase approximation (RPA) or quasiparticle RPA.
Different Skyrme-type interactions corresponding to incompressibilities in
the range 200 - 250 MeV are used. The results are discussed in comparison
with existing data on isoscalar giant dipole resonances. Two main issues are
raised, firstly the calculated giant resonance energies are somewhat higher
than the observed ones, and secondly a sizable fraction of strength is
predicted below 20 MeV which needs to be experimentally confirmed. 
\end{abstract}

\vspace{0.2cm}
 \par quasiparticle RPA. 
\vspace{0.5cm}

\newpage

The isoscalar giant dipole resonance (ISGDR) is a compressional mode, like
the well known isoscalar giant monopole resonance (ISGMR), and its energy is
related 
to the nuclear incompressibility $K_{\infty}$\cite{Stri}. 
For this reason it has been studied both experimentally
and theoretically for many years, as reviewed in \cite{Gar99}. 
It can be associated with the
operator
\begin{equation}
 \hat D = \sum_{i=1}^A r^3_i Y_{1\mu}(\hat r_i)
\label{dipole}\end{equation}
and it can be viewed as a non-isotropic compression mode.

Although some first indications about the energy location of this mode date
back to the beginning of the eighties, more recent evidence of the
ISGDR in $^{208}$Pb has been reported  \cite{Dav} from the 0$^0$
measurements  of 200 MeV inelastic $\alpha$-scattering at the Indiana
University cyclotron facility.
Further evidence based on extensive angular
distributions at near-0$^0$ angles has since come from 240 MeV inelastic
$\alpha$-scattering experiments at Texas A\&M University. The ISGDR strength
has been extracted in a large number of nuclei using a multipole decomposition
of the observed inelastic scattering spectra. The results for medium- and
heavy-mass nuclei like $^{90}$Zr, $^{116}$Sn, $^{144}$Sm and $^{208}$Pb
are reported in \cite{Clar}. The 
measured strength is spread over a wide energy range
between 15 and 30 MeV and it is claimed to exhaust nearly 100\% of the 
appropriate energy-weighted sum rule (EWSR).
The values of the centroid energy E$_0\equiv m_1/m_0$ are 26.3(4), 24.3(3),
23.0(3) and 20.3(2) MeV respectively in the four nuclei quoted above
(the moments $m_k$ of the strength distribution are defined as $m_k\equiv
\sum_n |<n|\hat D|0>|^2(E_n-E_0)^k$).

Here, we report the ISGDR results 
calculated with 
effective Skyrme interactions, within self-consistent Hartree-Fock (HF) 
plus Random Phase Approximation (RPA) in the case of $^{208}$Pb, 
and Hartree-Fock-BCS 
(HF-BCS) plus quasi-particle RPA (QRPA) for the other, non 
double-magic nuclei. 
The Skyrme forces used in
this work are: SkP~\cite{Dob84}, SGII~\cite{Ngu81}, SKM$^*$~\cite{Bar82},
SLy4~\cite{Cha98} and SkI2~\cite{Rei95}. They span a 
range of values of $K_\infty$ from 200 to 250 MeV.
The HF mean field is first 
calculated in coordinate space, then the single-particle spectrum of 
occupied and unoccupied states is built 
by diagonalizing the mean field on a harmonic
oscillator basis. Details of RPA calculations can be found in 
Ref.~\cite{PRC}. The dimension of the 1particle-1hole (1p-1h) space 
is fixed by requiring the exhaustion of the RPA $m_1$ sum rule. In the HF-BCS
calculations constant pairing gaps $\Delta$ are introduced according to the 
usual 12 MeV/$\sqrt{A}$ parametrization. The QRPA matrix equations are solved with 
a procedure which parallels what has been said for RPA, 
with the two quasi-particle
configurations replacing the 1p-1h ones. The method 
is the same as that of Ref.~\cite{Khan}. 

In the ISGDR problem, one has to face the question of the spurious state
associated with the center-of-mass motion which carries the same quantum
numbers $J^{\pi}= 1^{-}$. In a bona fide self-consistent RPA the spurious
state would appear as an eigenstate at zero energy, exhausting the whole
strength of the operator 
\begin{equation}
 \hat S = \sum_{i=1}^A r_i Y_{1\mu}(\hat r_i)
\label{spr}\end{equation}
and orthogonal to all other physical states. However, in actual calculations
the spurious state  is at low but not zero energy because of small numerical
inaccuracies 
and therefore, strength
associated with the operator $\hat S$ will be shared  among the physical
states.
Starting from the actual RPA set of states $|n^\prime\rangle$, we construct 
a new set of normalized states $|n\rangle$,
\begin{equation}
 |n\rangle = {\cal N}_n (|n^\prime\rangle - \alpha_n |S\rangle),
\label{new_n}\end{equation}
where the state $|S\rangle$ 
is defined as
\begin{equation}
 |S\rangle \equiv \hat S |0\rangle,
\end{equation}
$|0\rangle$ being the RPA vacuum. According to \cite{Geor} we associate to 
$|S\rangle$ the transition density
\begin{equation}
 \alpha_S {d\varrho_0\over dr}
\end{equation}
where $\varrho_0$ is the HF ground state density. 
The state $|n\rangle$ is required to satisfy the
condition $\langle n | \hat S | 0\rangle =0$, i.e., 
\begin{equation}
 \int dr r^3 (\delta\varrho_{n^\prime} - a_n {d\varrho_0\over dr}) = 0,
\label{expl_proj}\end{equation} 
where $\delta\varrho_{n^\prime}$ is the transition density of the RPA state
$|n^\prime\rangle$ defined in the usual way. The problem of the spurious state
normalization $\alpha_S$ is circumvented by the use of Eq.~(\ref{expl_proj}) 
since $a_n\equiv \alpha_n\alpha_S$ is well behaved (i.e., not divergent). 

The difference between the strength distributions associated with the states 
$|n^\prime\rangle$  and $|n\rangle$ is shown in the top-left corner of 
Fig.~1, for the typical case of $^{208}$Pb with the force SGII.
The strengths are essentially the same 
in the energy range which will be denoted ``giant resonance (GR) region'' 
(this range is evident from the plot but it is explicitly indicated in 
Tables~1 and \ref{table_all_nuclei}). At lower energies, 
omitting the projection procedure can lead to
a serious overestimation of the ISGDR strength. It is clear nevertheless 
from Fig.~1 that a non-negligible amount of 
non-spurious strength is present 
in the energy range which will be called ``low-energy region''. 
This low-lying strength is due to 1
$\hbar\omega$ excitations, which of course can contain strength associated
with the $\hat D$ operator.

Another way of eliminating the spurious strength is to
keep the $|n^{\prime} \rangle$ states and to replace the 
operator (\ref{dipole}) by 
\begin{equation}
 \hat D_{modif} = \sum_{i=1}^A (r^3_i-\eta r_i) Y_{1\mu}(\hat r_i),
\end{equation}
where $\eta={5\over 3}\langle r^2 \rangle$. 
This prescription was derived and used in Ref.~\cite{Ngu81b}. 
Although the derivation is based on hydrodynamical-type arguments, 
one thus obtains strength distributions which are almost indistinguishable to 
those calculated with
the present projection procedure. One may also note that in 
Ref.~\cite{Ham98} a different prescription was used for the subtraction of
spurious strength resulting in  an almost disappearance of strength in the
low-energy region. 

In Fig.~1 we also show center-of-mass corrected 
strength distributions for the
other nuclei calculated with a typical interaction, namely SGII.  
The general features are: a) a large fraction of the
strength lies in the GR region, and 
b) a non negligible
amount of strength is in the low-energy region. The latter region contains  
about 20\% of the ISGDR energy-weighted sum rule. These features are
common to the results obtained with the other interactions. A more detailed
analysis in terms of the moments $m_0$ and $m_1$ is reported 
in Table~\ref{table_ISGDR_manyforc} 
for $^{208}$Pb and all interactions, and in Table~\ref{table_all_nuclei}
for all 4 nuclei and the SGII interaction.

In comparison with the existing data, there are two main issues to be faced.
First, there is a large discrepancy between predicted and measured GR
energies, much larger than in all other GR cases. 
This is the more puzzling that the same
model employed here was used successfully to describe the ISGMR in 
$^{208}$Pb~\cite{Gian}. Second, the calculations predict a sizeable amount of
strength at low-energy, which needs to be experimentally 
confirmed~\cite{Youn}. These features are common to the calculated strength 
distributions of the operator $j_1(qr) Y_{1\mu}(\hat r)$, which is a 
generalization of Eq.(1). In particular, they remain peaked at the same 
energies as the strength distribution 
of $\hat D$ for values of $q$ up to 0.6 fm$^{-1}$. 

In
Fig.~2 we show the predicted peak and
centroid energies of the
GR region for various nuclei as a function of $K_{\infty}$.
The experimental values of $E_0$ for the GR region quoted above~\cite{Clar} would be
outside the figure, except for $^{90}$Zr. 
The discrepancy appears very severe in Pb and Sn. In what follows we
concentrate on Pb because it is the nucleus where the HF+RPA model should
work better.

Earlier RPA calculations~\cite{Gogn} performed with the
finite range Gogny interaction
already found that the ISGDR energy was in the range of 26 MeV,
in qualitative agreement with the present results and with
Ref.~\cite{Ngu81b}.
One might expect that effects beyond RPA, like the coupling to
2p-2h excitations would somewhat lower the centroid energy. However, the
calculations of Ref.~\cite{Colo} find a downward shift of
less than 1 MeV.
The ISGDR has also been calculated in the relativistic RPA
approach~\cite{Giai} in $^{208}$Pb and $^{144}$Sm
and it is found that, for 
effective lagrangian 
parametrizations 
corresponding to $K_{\infty}$ in the range 200-270 MeV the energy of the ISGDR
is of the order of 25 MeV.
Thus, the question of understanding the observed values of $E_0$ is still open.

As for the low-energy region,
our analysis of the configurations involved for instance in $^{208}$Pb,
shows that the strength comes from bound-to-bound neutron transitions like
$h_{9/2}\rightarrow i_{11/2}$, $i_{13/2}\rightarrow j_{15/2}$ and, in some
cases, $f_{5/2}\rightarrow g_{7/2}$. In the data reported
in Ref.~\cite{Clar} no low-lying strength is present. Further analysis of
the same data is currently in progress~\cite{Youn},  which may reveal the
presence of isoscalar strength around the region of the isovector dipole.

In conclusion, we report in this paper HF+RPA and HF-BCS+QRPA 
calculations of the ISGDR in $^{90}$Zr,
$^{112}$Sn, $^{144}$Sm and $^{208}$Pb nuclei. Two general features appear
from the calculated strength distributions: some large resonance-type
distribution of strength in the 110$A^{-1/3}$ MeV energy region and some
smaller, but still sizeable fraction of the strength below 20 MeV. These
two characteristic features do not seem to agree quantitatively with the
observation.

We thank D.H. Youngblood for helpful discussions about the experimental data 
analysis, and U. Garg for discussions. 
P.F.B. thanks IPN-Orsay for the warm hospitality during the time when this
work was completed.

\newpage

\begin{figure}
Fig. 1. ISGDR strength distributions calculated with the interaction SGII 
and corrected for center-of-mass effects. 
In the case of $^{208}$Pb 
the dashed line corresponds
to a calculation without proper subtraction of the spurious 
center-of-mass state (see text).
\end{figure}

\begin{figure}
Fig. 2. Centroid energies $E_0$ (black circles) and peak energies (open
circles) of the GR regions, determined by using
different Skyrme interactions 
and plotted as a function of $K_\infty$. The full and dashed
lines are drawn as a guide to the eye. 
\end{figure}

\newpage

\begin{table}[t]
\newlength{\digitwidth} \settowidth{\digitwidth}{\rm 0}
\catcode`?=\active \def?{\kern\digitwidth}
\caption{
Integral properties of
the ISGDR strength distribution in $^{208}$Pb calculated with different
forces. The corresponding incompressibilities are indicated in MeV. Results
for the low-energy region (0-17 MeV) and the GR
region (17-30 MeV) are separated. For each region the 
values of $m_0$ and $m_1$ are given in units 10$^5$fm$^6$ and 
10$^6$fm$^6\cdot$MeV, respectively. The last column shows (in MeV) the
centroid energies $m_1/m_0$ of the GR region, and the peak energies in
parenthesis. }
\vspace{0.5cm}
\label{table_ISGDR_manyforc}
\begin{tabular*}{\textwidth}{@{}|l@{\extracolsep{\fill}}c|cc|ccc|}
\cline{1-7}
                 Force & $K_\infty$  
                 & \multicolumn{2}{c|}{Low-energy region }
                 & \multicolumn{3}{c|}{GR region }
                 \\
                 && $m_0$ & $m_1$ & $m_0$ & $m_1$ &
                 $m_1/m_0$ 
                 \\
\hline
SkP              & 201  & 0.85   & 0.89  & 1.86  & 4.25  & 22.8 (23.6)  \\
SGII             & 215  & 0.90   & 0.98  & 1.61  & 3.85  & 23.9 (24.1)  \\
SkM$^*$          & 217  & 0.94   & 1.01  & 1.67  & 3.96  & 23.7 (24.2)  \\
SLy4             & 230  & 0.92   & 0.99  & 1.51  & 3.67  & 24.3 (25.2)  \\
SkI2             & 241  & 1.02   & 1.05  & 1.58  & 3.88  & 24.6 (25.3)  \\ 
\hline
\end{tabular*}
\end{table}

\begin{table}[b]
\caption{
Same as in Table~1, for the 4 nuclei calculated with SGII.
}
\vspace{0.5cm}
\label{table_all_nuclei}
\begin{tabular*}{\textwidth}{@{}|r@{\extracolsep{\fill}}|c|cc|ccc|}
\hline
                 Nucleus & Range of &  
                 \multicolumn{2}{c|}{Low-energy region}
                 & \multicolumn{3}{c|}{GR region}
                 \\
                 & GR region & $m_0$ & $m_1$ & $m_0$ & $m_1$ &
                 $m_1/m_0$ 
                 \\
\hline
$^{208}$Pb  & 17-30     & 0.90   & 0.98  & 1.61  & 3.85  & 23.9 (24.1)  \\
$^{144}$Sm  & 21-32     & 0.34   & 0.46  & 0.59  & 1.57  & 26.6 (26.5)  \\
$^{116}$Sn  & 20-35     & 0.20   & 0.25  & 0.40  & 1.10  & 27.5 (28.4)  \\
$^{90}$Zr   & 22-40     & 0.11   & 0.16  & 0.21  & 0.63  & 30.0 (26.7,32.5)   \\ 
\hline
\end{tabular*}
\end{table}

\end{document}